\documentclass[options]{JHEP3}
\title{\bf Poisson-Lie T-dual sigma models on supermanifolds}
\author {  A. Eghbali, A. Rezaei-Aghdam  \\Department of Physics, Faculty of science,
Azarbaijan University of Tarbiat Moallem , 53714-161, Tabriz,
Iran  \\ E-mail: \email{a.eghbali@azaruniv.edu,\\ Corresponding
Author: rezaei-a@azaruniv.edu }}

\abstract{We investigate Poisson-Lie symmetry for T-dual sigma
models on supermanifolds in general and on Lie supergroups in
particular. We show that the integrability condition on this super
Poisson-Lie symmetry is equivalent to the super Jacobi identities
of the Lie super-bialgebras. As examples we consider models
related to  four dimensional Lie super-bialgebras ${\it
((2A_{1,1}+ 2A)^1, {D^{10}}_{\hspace{-3mm}p=\frac{1}{2}}})$ and
${\it ((2A_{1,1}+ 2A)^1, I})$. Then generally it is shown that for
Abelian case  $({\bf g} , I)$ the super Poisson-Lie T-duality
transforms the role of fermionic (bosonic) fields in the model to
bosonic (fermionic) fields on the dual model and vice versa. }

\begin{document}
\section {\bf Introduction }
Field theories with supermanifold as target space  have recently
received considerable attention, because of their interesting
applications in both string theory and condensed matter physics.
$WZNW$ models on supergroups are related to local logarithmic
conformal field theories  \cite{{scho},{Gotz},{Saleur}} .
Moreover, sigma model on Lie supergroups provides the building
blocks of string theory in $AdS$ backgrounds; for example,
superstring theory on $AdS_3 \times S^3$ is related to a sigma
model on $PSU(1,1|2)$ \cite{{Berk},{Bersh},{Berkovits}}.
Furthermore, the study of two-dimensional topological sigma model
on supermanifolds gives a better understanding of mirror symmetry
and its relation to $T$-duality \cite{{seth},{Schwarz}}.
 $T$-duality is the most important symmetries of string theory
\cite{Giv}. On the other hand Poisson-Lie $T$-duality, a
generalization of $T$-duality, does not require existence of
isometry in the original target manifold (as in usual
$T$-duality) \cite{{K.S1},{K.S2},{K.S3},{JR}}; the integrability
of the Noether$^{,}$s currents associated with the action of group
$G$ on the target manifold is enough to have this symmetry. In
these models, the components of the Noether$^{,}$s currents play
the role of flat connection, i.e. they satisfy Maurer-Cartan
equations with group structure of $\tilde G$ (with the same
dimension of $G$) \cite{{K.S1},{K.S2}} so that $G$ and $\tilde G$
have Poisson-Lie structure and their Lie algebras form a Lie
bialgebra \cite{Derin}. In \cite{Parkh} it is shown that
Poisson-Lie $T$-duality in $(2,2)$ superconformal $WZNW$ and
Kazama-Suzuki models acts as a mirror symmetry. So the study of
Poisson-Lie $T$-dual sigma models on supermanifolds and its
relation to mirror symmetry is an interesting problem. However,
to our knowledge Poisson-Lie $T$-dual sigma models on
supermanifolds have not been studied until now. In this paper, we
investigate this problem.

The paper is organized as follows. In section two we generalize
the Poisson-Lie symmetry to sigma models on supermanifolds and
show that the, integrability condition on this super Poisson-Lie
symmetry gives the super Jacobi identities of Lie
super-bialgebras \cite{N.A}. The super Poisson-Lie $T$-dual sigma
models on supergroups are investigated in section three. As the
fourth section, we give  examples of four-dimensional Lie
super-bialgebras ${\it ((2A_{1,1}+ 2A)^1,
{D^{10}}_{\hspace{-3mm}p=\frac{1}{2}}})$ \cite{RE} and ${\it
((2A_{1,1}+ 2A)^1, I})$ \cite{R} and for the second example we
give canonical transformation that relate model and its dual one.
Then in section five,  we  show that in general for the Abelian
case $({\bf g} , I)$ the super Poisson-Lie $T$-duality transforms
the role of fermionic (bosonic) fields in the model to bosonic
(fermionic) fields on the dual model and vice versa. Some remarks
are discussed in as concluding section.

\section{\bf Super Poisson-Lie symmetry in sigma models on supermanifolds }
Consider two dimentional sigma models on supermanifolds
\footnote{Here we use the notation presented by DeWitt$^{,}$s in
\cite{D}. For example the transformation properties of upper and
lower right indices to the left one are as follows:

$$
^iT_{jl...}^{\;k}=T_{jl...}^{ik},\qquad
_jT^{ik}_{l...}=(-1)^j\;T_{jl...}^{ik},
$$
where indices on the exponent of $(-1)$ show the Grassmann degrees
of variables. }  $M$as target space with background matrix
${\hspace{-0.5mm}_\mu {\cal E}}\hspace{-0.5mm}_{ \nu}(x)
={\hspace{-0.5mm}_\mu G }\hspace{-0.5mm}_{
\nu}(x)+{\hspace{-0.5mm}_\mu B }\hspace{-0.5mm}_{ \nu}(x) $ as
function of coordinates $x^\mu$ \footnote {Here we consider
bosonic worldsheet with light cone coordinates: $\xi^
\pm=\frac{1}{2}(\tau \pm \sigma)$}
\begin{equation}
S\;=\;\frac{1}{2}\int\!d\xi^{+}\wedge
d\xi^{-}\;\partial_{+}x^\mu\; {\hspace{-0.5mm}_\mu {\cal
E}}\hspace{-0.5mm}_{ \nu}\;
\partial_{-}x^\nu  = \frac{1}{2}\int\!d\xi^{+}\wedge d\xi^{-} L.
\end{equation}
Suppose that a supergroup $G$ acts from right on $M$ freely. Then
with the use of left invariant supervector field
${{_iv}^\mu}^{(L,\;l)}$ (defined with left derivative)
\begin{equation}
{_iv}^{(L,\;l)} \;=\;{{_iv}^\mu}^{(L,\;l)}\;
\frac{{\overrightarrow{\partial}}}{\partial x^\mu}\;\qquad,
(i=1,...,dimG)
\end{equation}
corresponding to this action, one can compute variation of $S$
under transformation $x^\mu \longrightarrow x^\mu +
\epsilon^i(\xi^{+}, \xi^{-}) {_iv}^\mu$ as follows\footnote{From
now on we will omit the superscripts $(L,\;l)$ on ${{_iv}^\mu}$.
}:
\begin{equation}
\delta S \;=\;\frac{1}{2}\int\!d\xi^{+} \wedge
d\xi^{-}\;\varepsilon^i\;(-1)^{\lambda+i\lambda}\;\partial_{+}
x^\lambda\; {\cal L}_{_iv}\;{\cal E}_{\lambda \nu}\;\partial_{-}
x^\nu-\frac{1}{2}\int\!d\varepsilon^i \wedge \star {_iJ},
\end{equation}
where the Lie superderivative ${\cal L}_{_iv}\;{\cal E}_{\lambda
\nu}$ and Hodge star of Nother$^{,}$s current have the following
forms respectively
\begin{equation}
{\cal L}_{_iv}\;{\cal E}_{\mu
\nu}\;=\;(-1)^{i\mu+\mu+\lambda}\;{\overrightarrow {\partial_{\mu}
}}\;{_iv}^\lambda\; {\cal E}_{\lambda
\nu}+{_iv}^\lambda\;{\overrightarrow {\partial_{\lambda} }}\;{\cal
E}_{\mu \nu} +(-1)^{\mu \nu+\mu
\lambda+i\nu+\lambda+\nu}\;{\overrightarrow {\partial_{\nu}
}}\;{_iv}^\lambda\; {\cal E}_{\mu \lambda },
\end{equation}
\begin{equation}
\star {_iJ}\; =\;(-1)^{\mu+\lambda}\;{_iv}^\mu \;\partial_{+}
x^\lambda\;{\cal E}_{\lambda \mu}\;d\xi^{+}
-(-1)^{\mu}\;{_iv}^\mu\;{\cal E}_{\mu \nu}\;\partial_{-} x^\nu
d\xi^{-}.
\end{equation}
By direct calculation, one can consider
\begin{equation}
d \star {_iJ}\; =\; [(-1)^{\lambda+i\lambda}\;\partial_{+}
x^\lambda\;{\cal L}_{_iv}\;{\cal E}_{\lambda \nu}\;\partial_{-}
x^\nu +{_iv}^\mu\; (equations\; of\; motion)]\; d\xi^{-} \wedge
d\xi^{+},
\end{equation}
where the equations of motion have the following form
\bigskip
$$
-(-1)^{\lambda+\mu\lambda}\;\partial_{+} x^\lambda\;{\overrightarrow
{\partial_{\mu} }}\;{\cal E}_{\lambda \nu}\;
\partial_{-}x^\nu+(-1)^{\mu}\;\partial_{+} x^\lambda\;{\overrightarrow {\partial_{\lambda} }}\;{\cal E}_{\mu \nu}\;
\partial_{-} x^\nu+(-1)^{\mu}\;{\cal E}_{\mu \nu}\;\partial_{+}\partial_{-} x^\nu
$$
\begin{equation}
+(-1)^{\lambda+\mu}\;
\partial_{-}\partial_{+} x^\lambda\;{\cal E}_{\lambda \mu}
+(-1)^{\lambda+\mu}\;\partial_{+} x^\lambda\;
\partial_{-}x^\nu\;{\overrightarrow {\partial_{\nu} }}\;{\cal E}_{\lambda
\mu}=0.
\end{equation}
Thus, on extremal surface we have $\delta S =0$ and
\begin{equation}
d \star{_iJ}\; =\; -[(-1)^{\lambda+i\lambda}\;\partial_{+}
x^\lambda\;{\cal L}_{_iv}\;{\cal E}_{\lambda \nu}\;\partial_{-}
x^\nu ]\; d\xi^{+} \wedge d\xi^{-}.
\end{equation}
If ${\cal L}_{_iv}\;{\cal E}_{\lambda \nu}=0$, then $G$ is a
superisometry group of $M$ and we have conserved currents. On the
other hand, if $\star{_iJ}$ on extremal surfaces satisfy
Maurer-cartan equation \cite{D}
\begin{equation}
d \star{_iJ} \;= \;-(-1)^{jk}\;\frac{1}{2} {\tilde{f}^{jk}}_{\; \;
\; \; \; i} \star{_jJ} \wedge \star {_kJ},
\end{equation}
where, ${\tilde{f}^{jk}}_{\; \; \; \; \; i}$ are structure
constants of Lie superalgebras $\tilde {\bf g}$ (with the same
dimension of ${\bf g}$ (Lie superalgebra of $G$)); then the {\em
super Poisson-Lie symmetry} will be
\begin{equation}
{\cal L}_{v_i}({\cal E}_{\lambda \nu}) =(-1)^{\mu+\mu{'}+\lambda
\mu{'}+jk+\mu k+ \mu \mu{'}+\nu \lambda+ \mu{'} \nu}\;
{\tilde{f}^{jk}}_{\;  \; \; \; i}\;{_jv}^\mu\; {_kv}^\mu{'}
\;{\cal E}_{\mu \nu}\;{\cal E}_{\lambda \mu{'}},
\end{equation}
where this formula is a generalization of usual Poisson-Lie
symmetry \cite{K.S1} to sigma models on supermanifolds.\\
Now, using integrability condition on Lie superderivative
\begin{eqnarray}
{\cal L}_{[{_iv} , {_jv}]}({\cal E}_{\lambda \nu})\;=\;[{\cal
L}_{_iv} , {\cal L}_{_jv}]{\cal E}_{\lambda \nu}= {\cal
L}_{_iv}\;{\cal L}_{_jv}\;{\cal E}_{\lambda \nu}-(-1)^{ij}\;{\cal
L}_{_jv}{\cal L}_{_iv}\; {\cal E}_{\lambda \nu},
\end{eqnarray}
and after some computations, we see that the structure constants
of Lie superalgebras ${\bf g}$ and $\tilde {\bf g}$ must be
satisfied in the following relations
\begin{equation}
{f^k}_{ij}\;{\tilde{f}^{ml}}_{\;  \; \; \;
k}\;=\;(-1)^{il}\;{f^m}_{ik}\;{\tilde{f}^{kl}}_{\;  \; \; \; j}
+{f^l}_{ik}\;{\tilde{f}^{mk}}_{\;  \; \; \;
j}+{f^m}_{kj}\;{\tilde{f}^{kl}}_{\;  \; \; \; i}
+(-1)^{mj}\;{f^l}_{kj}\;{\tilde{f}^{mk}}_{\;  \; \; \; i},
\end{equation}
where these are the mixed superJacobi identities of the Lie
super-bialgebras $({\bf g},\tilde {\bf g})$ \cite{{R},{J.z}}.\\
In the same way, one can consider the dual sigma models with
background matrix ${\hspace{-0.5mm}_\mu {\tilde{\cal
E}}}\hspace{-0.5mm}_{ \nu}$; where the supergroup ${\tilde G}$
acts freely on $M$ and the roles of ${\bf g}$ and ${\tilde {\bf g}
}$ are exchanged.\\

\section{\bf Super Poisson-Lie T-dual sigma models on supergroups}

When the supergroup $G$ acts transitively and freely on $M$, the
target can be identified with the supergroup $G$. In this case,
in order to obtain T-dual sigma models, one can consider the
equation of motion for the action on the  Drinfeld superdouble $D$
\cite{K.S2}
\begin{equation}
<\partial_{\pm}l l^{-1} , \varepsilon^{\mp}> \;= \;0,
\end{equation}
where $l(\xi^+,\xi^-)$ is a map from world sheet to  Drinfeld
superdouble $D$ and $<.,.>$ is the invariant bilinear form on the
double and $\varepsilon^\mp$ are $n$ dimensional orthogonal super
vector spaces such that $\varepsilon^+ + \varepsilon^-$ spans Lie
superalgebra ${\cal D}= {\bf g} \bigoplus \tilde {\bf g}$. Now by
using decomposition $l$ in the vicinity of the unit element of
$D$ \cite{N.A}
\begin{equation}
l(\xi^{+} , \xi^{-})\; =\; g(\xi^{+} , \xi^{-}) \tilde{h}(\xi^{+}
, \xi^{-}),\qquad  (\; g\in G , \quad \tilde{h}\in \tilde{G}\; )
\end{equation}
we obtain from (3.1)
\begin{equation}
<g^{-1} \partial_{\pm}g + \partial_{\pm}\tilde{h}\hspace{0.5mm}
\tilde{h}^{-1} ,g^{-1} \varepsilon^{\mp} g> \;=\; 0 .\hspace{3cm}
\end{equation}
On the other hand, for the super vector spaces $\varepsilon^\mp$
we have
\begin{equation}
g^{-1} \varepsilon^{\pm} g\; = \;Span\{X_i \pm E^{\pm}_{ij}(g)
\tilde{X}^j\},
\end{equation}
where super matrices $E^{\pm}$ are supertranspose of each other
($E^{-}_{ij}=(-1)^{ij}E^{+}_{ji}$), $ \{X_i\}$ and
$\{\tilde{X}^i\}$ are bases of Lie superalgebras ${\bf g}$ and
$\tilde {\bf g}$ such that
$$
<X_i , X_j> \;= \;<\tilde{X}^i , \tilde{X}^j>\; = \;0,
$$
\begin{equation}
\qquad < X_i , {\tilde X}^j >\; =\; {\delta_ i} \hspace{1mm}^j
\;=\; (-1)^{ij}\; {\delta ^j} \hspace{1mm}_i \; =\; (-1)^{ij} <
{\tilde X}^j , X_i >,
\end{equation}
and we have\footnote{Here one must use of superdeterminant and
superinverse formula \cite{D}. }
\begin{equation}
E^{+}(g) = \Big(a(g) + E^{+}(e)\hspace{0.5mm}b(g)\Big)^{-1}
E^{+}(e)\hspace{0.5mm}d(g),
\end{equation}
such that
\begin{equation}
g^{-1} X_i\; g\; =\;{a(g)_i}\;^k\;{_k X}=(-1)^k\;{a(g)_i}\;^k
X_k,
\end{equation}
\begin{equation}
g^{-1} \tilde{X}^j g\; =\; {b(g)}^{jk}\;{_k X} + {d(g)^j}\;_k
\tilde{X}^k=(-1)^k\;{b(g)}^{jk}\;{X_k} + {d(g)^j}\;_k \tilde{X}^k,
\end{equation}
Now by using (3.3)-(3.5) we have
$$
{A_{+}}_i(g)\; :=\; (\partial_{+}\tilde{h}\hspace{0.5mm}
\tilde{h}^{-1})_i\; =\;(-1)^l\; (g^{-1} \partial_{+}g )^l
E^{+}_{li}(g), \\
$$
\begin{equation}
{A_{-}}_i(g) \;:= \;(\partial_{-}\tilde{h}\hspace{0.5mm}
\tilde{h}^{-1})_i\; =\; -E^{+}_{il}(g) (g^{-1} \partial_{-}g )^l,
\hspace{7mm}
\end{equation}
where $A_\pm$ are right invariant one forms on $\tilde{G}$ and
satisfy in the following flat connection relation
\begin{equation}
\partial_{+} A_{- i}(g) - \partial_{-} A_{+ i}(g) -(-1)^{j\hspace{0.5mm}l}\;
 {\tilde{f}^{j\hspace{0.5mm}l}}_{\; \; \; \; \; i}\; A_{- j}(g) A_{+ l}(g) =
 0.
\end{equation}
Indeed one can observe that the above equation results in the
equations of motions and super Poisson-Lie symmetry of the
following action
\begin{equation}
S\;=\;\frac{1}{2}\int\! \;(g^{-1} \partial_{+} g)^i\;{_i{
E^+_j}}(g) \; (g^{-1} \partial_{-} g)^j\; d\xi^{+} \; d\xi^{-}.
\end{equation}
To see this it is convenient to use the following definition for
the left invariant one forms with left derivative
\cite{{N.A},{egh}}
\begin{equation}
(g^{-1} \partial_{+} g)^i\;=\;{
L}\hspace{-0.5mm}^{(l)^i}_{+}\;=\;\partial_{+} x^\mu \; {_\mu
L}^{(l)^i},
\end{equation}
\begin{equation}
(g^{-1} \partial_{-} g)^j\;=\; {
L}\hspace{-0.5mm}^{(l)^j}_{-}\;=\;{{^jL}^{(l)^t}}_\nu \;
\partial_{-} x^\nu,
\end{equation}
where the superscript $t$ stands for supertransposition. On the
other hand, by use of $\;\;$ $< {_iv}\;,\; {
L}\hspace{-0.5mm}^{(l)j}
>\;=\; {_i\delta}^j $ with ${
L}\hspace{-0.5mm}^{(l)j}={\overrightarrow dx^\nu}\;{_\nu
L}\hspace{-0.5mm}^{(l)j}$ we have
\begin{equation}
{_iv}^\mu={_iL}\hspace{-0.5mm}^{(l){\mu^{\;{-1}}}}\hspace{-0.5mm}
\qquad, \qquad (-1)^{i+i\nu}\;{{^\nu
L}\hspace{-0.5mm}^{(l)}\hspace{-1mm}_i}\hspace{-0.55mm}^{\;{-t}}
={_iv}^\nu.
\end{equation}
Then from (3.9) we will have
$$
{A_{+}}_i(g) \; =\; (-1)^{i+i\nu}\; \partial_{+} x^\mu
\;{\hspace{-0.5mm}_\mu {\cal E}}\hspace{-0.5mm}_{ \nu}\;
{_iv}^\nu,\hspace{1.75cm}
$$
\begin{equation}
{A_{-}}_i(g) \;=\; -(-1)^i\;{_iv}^\mu\;{\hspace{-0.5mm}_\mu {\cal
E}}\hspace{-0.5mm}_{ \nu} \;\partial_{-} x^\nu,\hspace{2cm}
\end{equation}
where
\begin{equation}
{\hspace{-0.5mm}_\mu {\cal E}}\hspace{-0.5mm}_{ \nu}\;=\;{_\mu
L}^{(l)^i}\;{_i{ E^+_j}}(g) \; {{^jL}^{(l)^t}}_\nu.
\end{equation}
Now using the above relations in (3.10) we attain the proper
result. Note that by use of
\begin{equation}
(g^{-1} \partial_{\pm}g )\;=\;{
R}\hspace{-0.5mm}^{(l)^i}_{\pm}\;g^{-1}{_iX} g\;=\;{
R}\hspace{-0.5mm}^{(l)^i}_{\pm}\;{_i a}^{(l)^j}(g)\;{_jX} ,
\end{equation}
where
\begin{equation}
(\partial_{\pm}g g^{-1})^i\;=\;{ R}\hspace{-0.5mm}^{(l)^i}_{\pm},
\end{equation}
 we have
\begin{equation}
{ L}\hspace{-0.5mm}^{(l)^i}_{\pm}\;=\;{
R}\hspace{-0.5mm}^{(l)^j}_{\pm}\;{_j a}^{(l)^i}(g),
\end{equation}
and one can rewrite the action of (3.11) in the following form:
\begin{equation}
S\;=\;\frac{1}{2}\int\!\; (\partial_{+}g g^{-1})^i\;{_i{
\mathbb{E}^+_j}}(g) \; (
\partial_{-} g g^{-1})^j\; d\xi^{+} \; d\xi^{-},
\end{equation}
where
\begin{equation}
{_i{\mathbb{E}^+_j}}(g)\;=\;{{_i\Big({{E}^{+^{-1}}}\hspace{-2mm}(e)+\Pi(g)\Big)}_j}^{-1},
\end{equation}
and the super Poisson structure has the following form:
\begin{equation}
\Pi\;=\;b{a}^{-1}.
\end{equation}
In the same way one can obtain the following super Poisson-Lie
symmetric dual sigma model
\begin{equation}
\tilde{S}\; =\; \frac{1}{2}\int\!\; (\partial_{+}
\tilde{g}\tilde{g}^{-1} )_i \;
\tilde{\mathbb{E}}^{ij}(\tilde{g})\; {_j(\partial_{-} \tilde{g}
\tilde{g}^{-1} )}\; d\xi^{+} \; d\xi^{-},
\end{equation}
with
\begin{equation}
\tilde{\mathbb{E}}^{ij}(\tilde{g})\; =\;{\Big( {\tilde
E}^{+^{-1}}\hspace{-2mm}(\tilde e)+ \tilde {\Pi}(\tilde
g)\Big)^{ij}}^{-1}\;,
\end{equation}
where
\begin{equation}
E^{\pm}(e)\; \tilde{E}^{\pm}(\tilde{e})\; =\;
\tilde{E}^{\pm}(\tilde{e}) \;E^{\pm}(e) = I.
\end{equation}

\section{ Examples }
In this section we consider Poisson-Lie T-dual sigma models
related to four dimensional Lie super bialgebras ${\it ((2A_{1,1}+
2A)^1, {D^{10}}_{\hspace{-3mm}p=\frac{1}{2}}})$ and ${\it
((2A_{1,1}+ 2A)^1, I})$ \footnote { These Lie super-bialgebras
are obtained in \cite{RE} in the same way as \cite{R} . Note that
for the Lie superalgebras with odd numbers of fermionic
coordinates the metric tensor is singular \cite{D} hence we
consider four dimensional Lie superalgebras with two fermionic
coordinates as an example.}.

\subsection { \bf Case A }

For Lie super-bialgebra ${\it ((2A_{1,1}+ 2A)^1,
{D^{10}}_{\hspace{-3mm}p=\frac{1}{2}}})$, we have the following
nonzero (anti) commutation relations to the basis \footnote {Here
$\{X_1,X_2,{\tilde{X}}^1,{\tilde{X}}^2 \}$ and
$\{X_3,X_4,{\tilde{X}}^3 ,{\tilde{X}}^4\}$ are bosonic and
fermionic bases respectively \cite{B}.}  $\{X_1,X_2,X_3,X_4\}$ for
$(2A_{1,1}+ 2A)^1$ and
$\{{\tilde{X}}^1,{\tilde{X}}^2,{\tilde{X}}^3,{\tilde{X}}^4 \}$
for ${D^{10}}_{\hspace{-3mm}p=\frac{1}{2}}$
$$
\{X_3,X_3\}=X_1, \qquad ,\{X_4,X_4\}=X_2,\hspace{4cm}
$$
$$
[\tilde{X}^1,\tilde{X}^2]=\tilde{X}^2, \qquad
[\tilde{X}^1,\tilde{X}^3]=\frac{3}{2}\tilde{X}^3,\hspace{4cm}
$$
$$
[\tilde{X}^1,\tilde{X}^4]=\frac{1}{2}\tilde{X}^4, \qquad
[\tilde{X}^2,\tilde{X}^4]=\tilde{X}^3,\hspace{4cm}
$$
$$
\hspace{0.5cm}[X_2 , \tilde{X}^1] =X_2,  \qquad  [X_2 ,
\tilde{X}^2] =-X_1, \qquad [X_3 , \tilde{X}^1]
=\frac{3}{2}X_3-\tilde{X}^3,
$$
$$
[X_3 , \tilde{X}^2] =X_4,  \qquad [X_4 , \tilde{X}^1]
=\frac{1}{2}X_4,  \qquad [X_4 , \tilde{X}^2]
=-\tilde{X}^4,\hspace{0.5cm}
$$
\begin{equation}
\{X_3 , \tilde{X}^3\} =\frac{3}{2}X_1,  \qquad \{X_3 ,
\tilde{X}^4\} =X_2,  \qquad \{X_4 , \tilde{X}^4\} =\frac{1}{2}X_1.
\end{equation}
Now using (3.17) with the following representation for the Lie
supergroups ${\bf (2A_{1,1}+ 2A)^1}$ and ${\bf
D^{10}_{P=\frac{1}{2}}}$:
\begin{equation}
g = e^{x X_1}\;e^{y X_2}\;e^{\psi X_3}\;e^{\chi X_4}, \qquad
\tilde{g} = e^{\tilde{x} \tilde{X}^1}\;e^{\tilde{y}
\tilde{X}^2}\;e^{\tilde{\psi} \tilde{X}^3}e^{\tilde{\chi}
\tilde{X}^4},
\end{equation}
where $\{x,y,\tilde{x},\tilde{y}\}$ and
$\{\psi,\chi,\tilde{\psi},\tilde{\chi}\}$ are bosonic and
fermionic coordinates of the Lie supergroups ${\bf (2A_{1,1}+
2A)^1}$ and ${\bf D^{10}_{P=\frac{1}{2}}}$; we have
$$
{R_{\pm}}\hspace{-2mm}^{(l)^i}=\pmatrix{ \partial_{\pm}x
-\frac{\psi}{2}\;\partial_{\pm}\psi& \partial_{\pm}y
-\frac{\chi}{2}\;\partial_{\pm}\chi& -\partial_{\pm}\psi &
-\partial_{\pm}\chi },
$$
\vspace{-3mm}
\begin{equation}
\Pi^{ij}(g)=\pmatrix{0 & -y &
\frac{3\psi}{2} & \frac{\chi}{2} \cr y & 0 & 0 & \psi \cr
-\frac{3\psi}{2} & 0 & 0 & 0 \cr -\frac{\chi}{2} & -\psi & 0 & 0}.
\end{equation}
Then using (3.19) and choosing ${_i{ E^+_j}}(e)$ as
\begin{equation}
{_i{ E^+_j}}(e)=\pmatrix{ 1 & 0 & 0 & 0 \cr  0 & 1 & 0 & 0 \cr 0 &
0 & 0 & 1 \cr 0 & 0 & -1 & 0 },
\end{equation}
the following model is derived
\[ \begin{array}{lcl}
\vspace{3mm} S&=&\frac{1}{2}\int\;\frac{1}{2(1+y^2)}
\Big\{(2-\frac{3\psi
\chi}{1+y^2})(\partial_{+}x\;\partial_{-}x+y\;\partial_{+}x\;\partial_{-}y-y\;
\partial_{+}y\;\partial_{-}x)\hspace{5cm}\\
\vspace{3mm}
&+&(2+\frac{3y^2\psi
\chi}{1+y^2})\partial_{+}y\;\partial_{-}y-[\chi+(1+2y)\psi]\partial_{+}x\;\partial_{-}\psi
\hspace{1cm}\\
\vspace{3mm}
&-&[-\chi+(1+2y)\psi]\partial_{+}\psi\;\partial_{-}x
+(3\psi-y\chi)(\partial_{+}x\;
\partial_{-}\chi-\partial_{+}\chi\;\partial_{-}x)\\
\vspace{3mm}
&+&[y\chi+(y-2)\psi]\partial_{+}y\;
\partial_{-}\psi+[y\chi-(y-2)\psi]\partial_{+}\psi\;\partial_{-}y\\
\vspace{3mm}
&-&(\chi+3y\psi)(\partial_{+}y\;\partial_{-}\chi+\partial_{+}\chi\;\partial_{-}y)
+[2(1+y^2)+(5-y)\frac{\psi
\chi}{2}]\partial_{+}\psi\;\partial_{-}\chi
\end{array}
\]
\vspace{-4mm}
\begin{equation}
-\;(1+2y)\psi\chi\;\partial_{+}\psi\;\partial_{-}\psi-[2(1+y^2)+(1+y)\frac{\psi
\chi}{2}]\partial_{+}\chi\;\partial_{-}\psi \Big\}d\xi^{+}\;
d\xi^{-},\hspace{1.5cm}
\end{equation}
where the action have the following background matrices
\begin{equation}
{\hspace{-0.5mm}_\mu { G} }\hspace{-0.5mm}_{
\nu}=\frac{1}{2(1+y^2)}\pmatrix{ 2-\frac{3\psi \chi}{1+y^2} & 0 &
-\chi  & 3\psi-y\chi \cr  0 & 2+\frac{3y^2\psi \chi}{1+y^2} &
(y-2)\psi & 0 \cr -\chi & (y-2)\psi & 0 &
2(1+y^2)+\frac{3}{2}\psi\chi  \cr 3\psi-y\chi & 0 &
-2(1+y^2)-\frac{3}{2}\psi\chi  & 0 },
\end{equation}
\begin{equation}
{\hspace{-0.5mm}_\mu { B} }\hspace{-0.5mm}_{
\nu}=\frac{1}{2(1+y^2)}\pmatrix{ 0 & (2-\frac{3\psi \chi}{1+y^2})y
& -(1+2y)\psi  & 0 \cr -(2-\frac{3\psi \chi}{1+y^2})y & 0 & y\chi
& -(\chi+3y\psi)\cr (1+2y)\psi & -y\chi & -(1+2y)\psi \chi &
\frac{1}{2}(2-y)\psi\chi \cr 0 & (\chi+3y\psi) &
\frac{1}{2}(2-y)\psi\chi& 0 }.
\end{equation}
For the dual model with
$$
{{\tilde {R}^{(l)}} }\hspace{-2mm}_{\pm
i}=\pmatrix{\partial_{\pm}{\tilde x} & e^{\tilde
x}\;\partial_{\pm}{\tilde y} &  e^\frac{3{\tilde
x}}{2}\;(\partial_{\pm}{\tilde \psi}+{\tilde
y}\partial_{\pm}{\tilde \chi}) &e^\frac{{\tilde x}}{2}\;
\partial_{\pm}{\tilde \chi} },
$$
\vspace{-3mm}
\begin{equation}
{\tilde \Pi}_{ij}(\tilde g)=\pmatrix{0 & 0 & 0 & 0 \cr 0 & 0 & 0
& 0 \cr 0 & 0 & -\frac{1}{3}(\tilde y^3 e^{3\tilde x}+e^{3\tilde
x}-1) & -\frac{\tilde y^2}{2} e^{2\tilde x} \cr 0 & 0 &
-\frac{\tilde y^2}{2} e^{2\tilde x} & -\tilde y e^{\tilde
x}},\hspace{2cm}
\end{equation}
and
\begin{equation}
\tilde{E}^{ij}(\tilde{e})=\pmatrix{ 1 & 0 & 0 & 0 \cr  0 & 1 & 0
& 0 \cr 0 & 0 & 0 & -1 \cr 0 & 0 & 1 & 0 },
\end{equation}
we have
\[ \begin{array}{lcl}
\vspace{3mm} {\tilde S}&=&\frac{1}{2}\int \Big
\{\partial_{+}{\tilde x}\;\partial_{-}{\tilde x}+e^{2\tilde
x}\;\partial_{+}{\tilde y}\;\partial_{-}{\tilde
y}-\frac{1}{\lambda}[\tilde y e^{4\tilde x}\;\partial_{+} {\tilde
\psi}\;\partial_{-} {\tilde \psi}\\
\vspace{3mm} &+&(\frac{{\tilde y^2}}{2}e^{4\tilde x}-e^{2\tilde
x})\;\partial_{+} {\tilde \psi}\;\partial_{-} {\tilde \chi}+
(\frac{{\tilde y^2}}{2}e^{4\tilde x}+e^{2\tilde x})\;\partial_{+}
{\tilde \chi}\;\partial_{-} {\tilde \psi}
\end{array}
\]
\vspace{-4mm}
\begin{equation}
+\;\frac{e^{\tilde x}}{3}(\tilde y^3 e^{3\tilde x}+e^{3\tilde
x}-3)\;\partial_{+} {\tilde \chi}\;\partial_{-} {\tilde \chi}]
\Big\}d\xi^{+}\; d\xi^{-},\hspace{1cm}
\end{equation}
such that for the background matrices we have
\begin{equation}
{\hspace{-0.5mm}_\mu {\tilde G} }\hspace{-0.5mm}_{ \nu}=\pmatrix{
1 & 0 & 0 & 0 \cr  0 & e^{2\tilde x} & 0 & 0 \cr 0 & 0 & 0 &
\frac{e^{2\tilde x}}{\lambda}\cr 0 & 0 & -\frac{e^{2\tilde
x}}{\lambda} & 0 }, \quad {\hspace{-0.5mm}_\mu {\tilde B}
}\hspace{-0.5mm}_{ \nu}=\pmatrix{0 & 0 & 0 & 0 \cr  0 & 0 & 0 & 0
\cr 0 & 0 & \frac{-\tilde{y}e^{4\tilde x}}{\lambda} &
\frac{-{\tilde y}^2 e^{4\tilde x}}{2\lambda} \cr 0 & 0 &
\frac{-{\tilde y}^2 e^{4\tilde x}}{2\lambda}& \frac{-e^{\tilde
x}}{3\lambda}(\tilde y^3 e^{3\tilde x}+e^{3\tilde
x}-3)},\hspace{1cm}
\end{equation}
where
$$
\lambda=\frac{{\tilde y^4}}{12}e^{4\tilde x}+\frac{{\tilde
y}}{3}e^{4\tilde x}-\frac{{\tilde y}}{3}e^{\tilde x}+1.
$$

\bigskip

\subsection { \bf Case B }

For Lie super-bialgebra ${\it ((2A_{1,1}+ 2A)^1, I})$, where $I$
is $(2,2)$ (with two bosonic and fermionic bases) Abelian
superalgebra, we have the following nonzero (anti) commutation
relations to the basis $\{X_1,X_2,X_3,X_4\}$ for $(2A_{1,1}+
2A)^1$ and
$\{{\tilde{X}}^1,{\tilde{X}}^2,{\tilde{X}}^3,{\tilde{X}}^4 \}$
for $I$:
$$
\{X_3,X_3\}=X_1, \qquad \{X_4,X_4\}=X_2,\hspace{3cm}
$$
\begin{equation}
[X_3 , \tilde{X}^1] =-\tilde{X}^3,  \qquad [X_4 , \tilde{X}^2]
=-\tilde{X}^4.\hspace{2.5cm}
\end{equation}
Using the representation in (4.2) we have $\Pi^{ij}(g)=0$ and if
we  choose ${_i{ E^+_j}}(e)$ as (4.4) the action of the model
will have the following form
$$
S\;=\;\frac{1}{2}\int\!d\xi^{+}\; d\xi^{-}\; \Big\{
\partial_{+}x\;\partial_{-}x+\partial_{+}y\;\partial_{-}y+
\partial_{+}\psi\;\partial_{-}\chi-\partial_{+}\chi\;\partial_{-}\psi
\hspace{3cm}
$$
\begin{equation}
-\;\frac{\psi}{2}\;(\partial_{+}x\;\partial_{-}\psi
+\partial_{+}\psi
\;\partial_{-}x)-\frac{\chi}{2}\;(\partial_{+}y\;
\partial_{-}\chi+ \partial_{+}\chi\;\partial_{-}y)
\Big\},\hspace{2.25cm}
\end{equation}
with the background matrices as
\begin{equation}
{\hspace{-0.5mm}_\mu { G} }\hspace{-0.5mm}_{ \nu}=\pmatrix{ 1 & 0
& 0 & 0 \cr  0 & 1 & 0 & 0 \cr 0 & 0 & 0 & 1 \cr 0 & 0 & -1 & 0
}\quad , \quad {\hspace{-0.5mm}_\mu { B} }\hspace{-0.5mm}_{
\nu}=\pmatrix{0 & 0 & -\frac{\psi}{2} & 0 \cr  0 & 0 & 0 &
-\frac{\chi}{2} \cr \frac{\psi}{2} & 0 & 0 & 0 \cr 0 &
\frac{\chi}{2} & 0 & 0 }.
\end{equation}
For the dual model we find
\begin{equation}
{\tilde {R_{\pm}} \hspace{-1mm}^{(l)}} \hspace{-2mm}_{i}=\pmatrix{
\partial_{\pm}{\tilde x} & \partial_{\pm}{\tilde y} & \partial_{\pm}{\tilde \psi} & \partial_{\pm}{\tilde \chi}}
 \qquad, \qquad {\tilde
\Pi}_{ij}(\tilde g)=\pmatrix{ 0 & 0 & 0 & 0 \cr  0 & 0 & 0 & 0
\cr 0 & 0 & -\tilde x & 0 \cr 0 & 0 & 0 & -\tilde y },
\end{equation}
then we have
$$
\tilde S=\frac{1}{2}\int\!d\xi^{+}\; d\xi^{-}\; \Big \{
\partial_{+}{\tilde x}\;\partial_{-}{\tilde x}+\partial_{+}{\tilde y}\;\partial_{-}{\tilde
y}\hspace{6.24cm}
$$
\begin{equation}
+\;\frac{1}{\tilde{x} \tilde{y}+1}(\partial_{+} {\tilde
\psi}\;\partial_{-} {\tilde \chi}-\partial_{+} {\tilde
\chi}\;\partial_{-} {\tilde \psi}-\tilde y\;\partial_{+} {\tilde
\psi}\;\partial_{-} {\tilde \psi}-\tilde x\;\partial_{+} {\tilde
\chi}\;\partial_{-} {\tilde \chi})\Big \},\hspace{0.83cm}
\end{equation}
with the following background matrices
\begin{equation}
{\hspace{-0.5mm}_\mu {\tilde G} }\hspace{-0.5mm}_{ \nu}=\pmatrix{
1 & 0 & 0 & 0 \cr  0 & 1 & 0 & 0 \cr 0 & 0 & 0 &
\frac{1}{\tilde{x} \tilde{y}+1} \cr 0 & 0 & \frac{-1}{\tilde{x}
\tilde{y}+1} & 0 } \quad , \quad  {\hspace{-0.5mm}_\mu {\tilde B}
}\hspace{-0.5mm}_{ \nu}=\pmatrix{0 & 0 & 0 & 0 \cr  0 & 0 & 0 & 0
\cr 0 & 0 & \frac{-\tilde{y}}{\tilde{x} \tilde{y}+1} & 0 \cr 0 & 0
& 0 & \frac{-\tilde{x}}{\tilde{x} \tilde{y}+1} }.
\end{equation}
Note that the background matrices of the model depend only on the
fermionic fields $\{\psi,\chi\}$ and for the dual model the
matrices depend only on the bosonic fields $\{\tilde x,\tilde
y\}$. As such we see that in this case the super Poisson-Lie
T-duality (super non-Abelian duality) transforms the role of
fermionic fields in the model to bosonic fields on the dual model.
In the next section we generalize this feature for general Abelian
Lie super-bialgebra $(\bf g , I)$. But before that, let us
investigate the physical equivalence of the model and its dual in
the case B by use of canonical transformations. The generating
functional for this equivalence is
$$
F[x , \tilde x]\;=\;-\frac{1}{2}\int\;d\sigma\;{\tilde
x}_i\;{{R}^{(l)^i}}\hspace{-3mm}_\sigma \hspace{8cm}
$$
\vspace{-3mm}
\begin{equation}
\hspace{2.5cm}=\;-\frac{1}{2}\int\;d\sigma\;\{{\tilde
x}\;\partial_{\sigma}x-{\tilde
x}\;\frac{\psi}{2}\;\partial_{\sigma}\psi+{\tilde
y}\;\partial_{\sigma}y-{\tilde
y}\;\frac{\chi}{2}\;\partial_{\sigma}\chi-{\tilde \psi}\;
\partial_{\sigma}\psi-{\tilde \chi}\;\partial_{\sigma}\chi\},
\end{equation}
This generating functional  produces the following canonical
transformations
\begin{equation}
{p}_{i}\;=\;\frac{\overleftarrow{\delta}F}{\delta { x}^{i}},\qquad
\qquad {\tilde p}^{i}\;=\;-\frac{\overleftarrow{\delta}F}{\delta
{\tilde x}_{i}},
\end{equation}
i.e.
$$
{p}_{1}\;=\;\frac{1}{2}\partial_{\sigma}{\tilde
x},\hspace{3.8cm}\qquad \qquad {\tilde
p}^{1}\;=\;\frac{1}{2}\partial_{\sigma}{ x}-\frac{1}{4} \psi\;
\partial_{\sigma}\psi,
$$
\vspace{-4mm}
$$
{p}_{2}\;=\;\frac{1}{2}\partial_{\sigma}{\tilde
y},\hspace{3.8cm}\qquad \qquad {\tilde
p}^{2}\;=\;\frac{1}{2}\partial_{\sigma}{ y}-\frac{1}{4} \chi\;
\partial_{\sigma}\chi,
$$
\vspace{-4mm}
$$
{p}_{3}\;=\;-\frac{1}{2}{\tilde x}\;\partial_{\sigma}{\tilde
\psi}-\frac{1}{4} \psi\;\partial_{\sigma}{\tilde
x}-\frac{1}{2}\partial_{\sigma}{\tilde \psi},\qquad \qquad {\tilde
p}^{3}\;=\;\frac{1}{2}\partial_{\sigma}{\psi},\hspace{1.65cm}
$$
\vspace{-4mm}
\begin {equation}
{p}_{4}\;=\;-\frac{1}{2}{\tilde y}\;\partial_{\sigma}{\tilde
\chi}-\frac{1}{4} \chi\;\partial_{\sigma}{\tilde
y}-\frac{1}{2}\partial_{\sigma}{\tilde \chi},\qquad \qquad {\tilde
p}^{4}\;=\;\frac{1}{2}\partial_{\sigma}{\chi},\hspace{1.65cm}
\end {equation}
with these canonical transformations the Hamiltonian of the model

$$
{\cal H}\;=\;{p_1}^2+{p_2}^2+ \psi\; p_1\; p_4+\frac{1}{2}\psi \;
\chi\; p_1 \;p_2 + 2p_3\; p_4-\chi\; p_2\;
p_3+\frac{1}{4}\partial_{\sigma}x\;
\partial_{\sigma}x
$$
\vspace{-5mm}
\begin {equation}
+\;\frac{1}{4}\partial_{\sigma}y\; \partial_{\sigma}y+
\frac{1}{2}\partial_{\sigma} \psi\;\partial_{\sigma}
\chi-\frac{1}{4}\psi\;\partial_{\sigma}x\;\partial_{\sigma}
\psi-\frac{1}{4}
\chi\;\partial_{\sigma}y\;\partial_{\sigma}\chi,\hspace{1.5cm}
\end {equation}
with
\begin {equation}
{p}_{i}\;=\;\frac{\overleftarrow {\partial} L} {\partial
(\partial_{\tau}x^i)},
\end {equation}
is equal to the Hamiltonian of the dual model
$$
{\tilde{\cal H}}\;=\;({\tilde p^1})^2+({\tilde p^2})^2+ 2({\tilde
x} \;{\tilde y}+1) {\tilde p^3}\; {\tilde p^4} + {\tilde x}
\;{\tilde p^3}\; \partial_{\sigma}{\tilde \chi}-{\tilde y}
\;{\tilde p^4}\;
\partial_{\sigma}{\tilde \psi}
$$
\vspace{-5mm}
\begin {equation}
 + \;\frac{1}{2}\partial_{\sigma}\tilde{ \psi}\;\partial_{\sigma}
\tilde{ \chi}+\frac{1}{4}(\partial_{\sigma}{\tilde x}\;
\partial_{\sigma}{\tilde x}+\partial_{\sigma}{\tilde y}\;\partial_{\sigma}{\tilde y} ).\hspace{1.5cm}
\end {equation}
with
\begin {equation}
{\tilde p}^{i}\;=\;\frac{\overleftarrow {\partial} {\tilde L}}
{\partial (\partial_{\tau}{\tilde x}_i)}.
\end {equation}
Therefore the two models (4.13) and (4.16) are physically
equivalent.

\section{\bf Super non-Abelian duality }

Here we consider Abelian Lie super-bialgebras $({\bf g} , I)$
where ${\bf g}$ and $I$ have $(m , 2n)$ m bosonic and $2n$
fermionic generators. We consider the following three cases in
terms of commutation relations of Lie superalgebra ${\bf g}$:

\smallskip
a)$\;$ The commutation relations for the generators
$\{X_A\}=\{X_1,\cdots ,X_m,X_{m+1},\cdots ,X_{m+2n} \}$ of $\bf
g$ have the following form:
\begin{equation}
[X_i ,
X_{m+a}]\;=\;\sum_{b=1}^{2n}\;{f^{m+b}}\hspace{-2mm}_{i,m+a}\;X_{m+b},
\qquad i=1,\cdots , m, \quad a =1,\cdots, 2n,
\end{equation}
i.e. we have only  commutation relation of bosons with fermions.
Now by use of the following parameterizations for the Lie
supergroup G
\begin{equation}
g\;=\;e^{x^1 X_1} \cdots e^{x^m X_m}\;e^{\psi^1 X_{m+1}}\cdots
e^{\psi^{2n} X_{m+2n}},
\end{equation}
we have
\begin{equation}
\partial _{\pm} g g^{-1}=\sum_{i=1}^{m} \partial _{\pm}x^i X_i +\sum
_{a=1}^{2n}\; \sum _{k_1, k_2,\cdots k_{m+a}=1}^{m+2n}\partial
_{\pm}{\psi}^a{{(e^{-x^m \chi_m})}_{m+a}}^{\;k_1}\cdots {{(e^{-x^1
\chi_1})}_{k_m}}^{\;k_{m+1}}\; X_{k_{m+a}},
\end{equation}
where $\chi_m$ is adjoint representation of the bosonic bases
$X_m$. By comparison of this relation with (3.18) we see that the
${ R}\hspace{-0.5mm}^{(l)^i}_{\pm}$ are functions of bosonic
coordinates of Lie supergroup G i.e. $\{x^i\}$. On the other hand
as for case B of section 4 we have
\begin{equation}
b=\Pi= 0, \qquad  \qquad  E^+_{ij}(g)\;=\; E^+_{ij}(e).
\end{equation}
In this way by use of (3.19) we see that the background matrix
depends only on the bosonic coordinates of the Lie supergroup G.
Furthermore by using of the commutation relations of Lie
super-bialgebras for $({\bf g} , I)$ we have \cite{R}
\begin{equation}
[X_i , {\tilde
X}^{m+a}]\;=\;\sum_{b=1}^{2n}\;{f^{m+b}}\hspace{-2mm}_{m+b,i}\;{\tilde
X}^{m+b},
\end{equation}
\begin{equation}
[X_{m+a} , {\tilde
X}^{m+b}]\;=-\;\sum_{i=1}^{m}\;{f^{m+b}}\hspace{-2mm}_{i,m+a}\;{\tilde
X}^{i},
\end{equation}
then one can obtain
\begin{equation}
{\tilde g}^{-1} X_i\; {\tilde g}\; =\;X_i+\sum _{a=1}^{2n}\;\sum
_{b=1}^{2n}\;{\tilde
\psi}_a\;{f^{m+a}}\hspace{-2mm}_{m+b,i}\;{\tilde X}^{m+b},
\end{equation}
\begin{equation}
{\tilde g}^{-1} X_{m+a}\; {\tilde g}\; =\;X_{m+a}-\sum
_{i=1}^{m}\;\sum _{b=1}^{2n}\;{\tilde
\psi}_b\;{f^{m+b}}\hspace{-2mm}_{m+a,i}\;{\tilde X}^{i},
\end{equation}
and by comparison of dual version of (3.8) we see that the matrix
$\tilde b$ depends only on the fermionic coordinates of $\tilde
G$; then ${\tilde \Pi}\;=\;\tilde b\; {\tilde a}^{-1}$ is only the
function of fermionic coordinates of $\tilde G$ and background
matrix of dual model depends only on the fermionic coordinates. In
this case super Poisson Lie T-duality transforms  the role of
bosonic fields in the model to the  fermionic fields on the dual
model.

\smallskip
b)$\;$ The commutation relations for the generators $\{X_A\}$ of
$\bf g$ have the following form
\begin{equation}
[X_{m+a} ,
X_{m+b}]\;=\sum_{i=1}^{m}\;{f^{i}}\hspace{-1mm}_{m+a,m+b}\;X_{i},
\end{equation}
i.e. we have only commutation relations of fermions with fermions.
In this case we have
\begin{equation}
\partial _{\pm} g g^{-1}\;=\;\sum_{i=1}^{m} \partial _{\pm}x^i X_i +\sum
_{a=1}^{2n} \partial _{\pm}{\psi}^a\;
X_{m+a}-\sum_{a>b=1}^{2n}\;\sum_{i=1}^{m}\;\partial
_{\pm}{\psi}^a\;{\psi}^b\;{f^{i}}\hspace{-1mm}_{m+a,m+b}\;X_{i},
\end{equation}
then the background matrix of the model depends only on the
fermionic fields. On the other hand, for the dual model we have
\begin{equation}
[X_{m+a} , {\tilde X}^i]\;=\;-\sum_{b=1}^{2n}\;
{f^{i}}\hspace{-1mm}_{m+b,m+a}\;{\tilde X}^{m+b},
\end{equation}
\begin{equation}
{\tilde g}^{-1} X_i\; {\tilde g}\; =\;X_i,\qquad \qquad {\tilde
g}^{-1} X_{m+a}\; {\tilde g}\;
=X_{m+a}+\sum_{i=1}^{m}\;\sum_{b=1}^{2n}\; {\tilde
x}_i\;{f^{i}}\hspace{-1mm}_{m+b,m+a}\; {\tilde X}^{m+b},
\end{equation}
then the matrix $\tilde b$ and background matrix of the dual model
depend only on the bosonic fields. In this case super Poisson Lie
T-duality transforms  the role of fermionic fields in the model to
the bosonic fields on the dual model (such as case B of section
4).

\smallskip
c)$\;$For the case
\begin{equation}
[X_i , X_j]\;=\sum_{k=1}^{m}\;\; {f^{k}}\hspace{-1mm}_{ij}\;X_{k},
\end{equation}
we have only commutation relations of bosons with bosons; the
background matrices of models and its dual are functions of
bosonic fields and super Poisson Lie T-duality transforms  the
bosonic fields  to the bosonics  ones (such as ordinary Poisson
Lie T-duality).

\section{\bf Conclusion }
We investigated Poisson-Lie $T$-duality for sigma models on
supermanifolds, especially on Lie supergroups. We show that for
the Abelian case $({\bf g} , I)$ the super Poisson Lie T-duality
transforms  the role of fermionic (bosonic) fields in the model to
bosonic (fermionic) fields on the dual model and vice versa. We
hope that the relationship between $T$-duality and mirror symmetry
will be better understood in this way. Furthermore, one can
investigate Poisson-Lie $T$-dual sigma models on supergroups
having conformal symmetry, as well as its relations to
superstring theories on $AdS$ backgrounds.  The study of
Poisson-Lie $T$-dual sigma models on low dimensional supergroups
\cite{{R},{RE}} with spectator fields and its relations to models
such as $2+1$ dimensional string cosmology coupled with fermionic
matter can be considered an other open problems, some of which
are under investigation.

\acknowledgments We would like to thank Sh. Moghadassi for
carefully reading the manuscript and useful comments.


\begin{thebibliography}{99}
\bibitem{scho}V. Schomerus and H. Saleur, {\it The $GL(1|1)$ $WZW$ model: From supergeometry to logarithmic
CFT }, \npb{734}{2006}{221-245}, \hepth{0510032}.
\bibitem{Gotz}
G.Gotz, T.Quella and V.Schomerus, {\it The WZW model on
$PSU(1,1|2)$}, \jhep{0703}{2007}{003}, \hepth{0610070}.
\bibitem{Saleur}
H. Saleur and V. Schomerus, {\it On the $SU(2|1)$ $WZNW$ model and
its statistical mechanics applications},
\npb{775}{2007}{312-340}, \hepth{0611147}.
\bibitem{Berk} N.Berkovits, C.Vafa, E.Witten,
 {\it Conformal Field Theory of Ads Backgrounds with Ramond-Ramond
Flux}, \jhep{9903}{1999}{018}, \hepth{9902098}.
\bibitem{Bersh} M. Bershadsky, S. Zhukov and A. Vaintrob, {\it $PSL(n|n)$ sigma
models as a conformal field theory }, \npb{559}{1999}{205-234},
\hepth{9902180}.
\bibitem{Berkovits} N. Berkovits, M. Bershadsky, T. Hauer, S. Zhukov and B. Zwiebach,
 {\it Superstring Theory on $AdS_2 \times S_2$ as a Coset
Supermanifold}, \npb{567}{2000}{61-86}, \hepth{9907200}.
\bibitem{seth}S. Sethi, {\it Supermanifolds, Rigid Manifolds and Mirror
Symmetry}, \npb{430}{1994}{31-50}, \hepth{9404186}.
\bibitem{Schwarz} A. Schwarz, {\it Sigma models having supermanifolds as target
spaces}, \lmp{38}{1996}{91-96}, \hepth{9506070}.
\bibitem{Giv} A. Giveon, M. Porrati, E.Rabinovici, {\it Target
Space Duality in String Theory}, \prep{244}{1994}{77-202},
\hepth{9401139}.
\bibitem{K.S1} C. Klim\v {c}ik and P. \v {S}evera, {\it Dual non-Abelian duality and
the Drinfeld double}, \plb{351}{1995}{445-462}.
\bibitem{K.S2} C. Klim\v {c}ik, {\it Poisson-Lie $T$-duality}, \npps{46}{1996}{116-121}, \hepth{9509095}.
\bibitem{K.S3} C. Klim\v {c}ik and P. \v {S}evera, {\it Poisson-Lie $T$-duality and
loop groups Drinfeld doubles}, \plb{372}{1996}{65-71},
\hepth{9512040}.
\bibitem{JR}M.A. Jafarizadeh and A. Rezaei-Aghdam, {\it Poisson-Lie
T-duality and Bianchi type algebras}, \plb{458}{1999}{477-490},
\hepth{9903152}.
\bibitem{Derin} V. G. Derinfeld, {\it Quantum groups} in:
Proc.ICM. MSRI, Berkeley. 1986. p. 798.
\bibitem{Parkh} S. G. Parkhomenko, {\it Mirror symmetry as a
Poisson-Lie T-dulity}, \mpla{13}{1998}{1041-1054},
\hepth{9710037}.
\bibitem{D}B. DeWitt, {\it Supermanifold}, Cambridge University Press 1992.
\bibitem {N.A} N. Andruskiewitsch,
\newblock  {\it Lie Superbialgebras and Poisson-Lie supergroups},
\newblock Abh. Math. Sem. Univ. Hamburg  63 (1993). 147-163.

\bibitem{R} A. Eghbali, F. Heidarpour and A. Rezaei-Aghdam, {\it Classification of two and
three dimensional Lie super-bialgebras}, {\tt arXiv:0901.4471
[math-ph]}.
\bibitem{RE} A. Eghbali and A. Rezaei-Aghdam, {\it Classification
of four-dimensional Lie super-bialgebras of type $(2 , 2)$}, work
in progress.
\bibitem {egh}  A. Eghbali and A. Rezaei-Aghdam, {\it Classical $r$-matrices of
two and three dimensional Lie super-bialgebras and their
Poisson-Lie supergroups}, {\tt arXiv:0908.2182 [math-ph]}.
\bibitem {J.z} C. Juszczak and J. T. Sobczyk, {\it Classification of low dimentional Lie
super-bialgebras}, \jmp{39}{1998}{4982-4992}.
\bibitem {B} N. Backhouse, {\it A classification of four-dimensional Lie
superalgebras}, \jmp{19}{1978}{2400-2402}.
\end{thebibliography}
\end{document}